# The Technical Debt Dataset

Valentina Lenarduzzi
Tampere University, Finland
valentina.lenarduzzi@tuni.fi

Nyyti Saarimäki
Tampere University, Finland
nyyti.saarimaki@tuni.fi

Davide Taibi
Tampere University, Finland
davide.taibi@tuni.fi

## ABSTRACT

Technical Debt analysis is increasing in popularity as nowadays researchers and industry are adopting various tools for static code analysis to evaluate the quality of their code. Despite this, empirical studies on software projects are expensive because of the time needed to analyze the projects. In addition, the results are difficult to compare as studies commonly consider different projects. In this work, we propose the Technical Debt Dataset, a curated set of project measurement data from 33 Java projects from the Apache Software Foundation. In the Technical Debt Dataset, we analyzed all commits from separately defined time frames with SonarQube to collect Technical Debt information and with Ptidej to detect code smells. Moreover, we extracted all available commit information from the git logs, the refactoring applied with Refactoring Miner, and fault information reported in the issue trackers (Jira). Using this information, we executed the SZZ algorithm to identify the fault-inducing and -fixing commits. We analyzed 78K commits from the selected 33 projects, detecting 1.8M SonarQube issues, 38K code smells, 28K faults and 57K refactorings. The project analysis took more than 200 days. In this paper, we describe the data retrieval pipeline together with the tools used for the analysis. The dataset is made available through CSV files and an SQLite database to facilitate queries on the data. The Technical Debt Dataset aims to open up diverse opportunities for Technical Debt research, enabling researchers to compare results on common projects.

## REFERENCE

## 1 INTRODUCTION

Companies commonly invest effort to improve the quality of their software by removing technical issues believed to impact software quality. Technical issues include any kind of information that can be derived from the source code or from the software process, such as usage of specific patterns, compliance with coding or documentation conventions, or architectural issues. If such issues are not fixed, they generate Technical Debt.

Technical Debt (TD) is a metaphor from the economic domain that refers to different software maintenance activities that are postponed in favor of the development of new features in order to get short-term payoff [5]. Just as in the case of financial debt, the additional cost will be paid later. The growth of TD commonly slows down the development process [5], [25].

Several Technical Debt measurement tools are available on the market (e.g., Better Code Hub[1], Coverity Scan[2], and SonarQube[3]). SonarQube is one of the tools most frequently adopted in research and industry [23] and it is adopted by more than 120K users worldwide[3] to improve the maintaineance of their systems [22].

Researchers have investigated several aspects of TD, trying to understand whether some specific issues can be considered harmful in certain contexts. For example, different research groups have investigated whether the presence of code smells increases change- or fault-proneness [29], [7], [14], [31], while other studies have investigated whether other code smells can impact maintenance effort [26], [38]. However, each research work is based on custom-built datasets, which are often not shared publicly or, in some cases, are not available anymore. As a result, the outcomes are not directly comparable, as they stem from different projects and contexts.

Moreover, in recent years, researchers have mainly focused their attention on code smells [10] and architectural smells [12]. However, the harmfulness of other technical issues detected by the tools commonly used in industry has not been investigated extensively. This is mainly due to the effort required to collect and analyze data with these tools.

To the best of our knowledge, only a few datasets provided TD data on open-source projects [32], [40] and a dataset containing fault information from Jira [30]. However, no dataset collects and shares information from different tools.

For this purpose, we have created the Technical Debt Dataset [36], which is a curated dataset containing measurement data from five tools executed on all commits of 33 projects from the Apache Software Foundation during the time frames reported in Table 1. The aim of this dataset is to enable researchers to work on a common set of data and thus compare their results.

The dataset was built by extracting the projects' data and analyzing it using several tools. To get the data, the projects' GitHub repositories were cloned, commit information was collected from the git log using PyDriller [39], refactorings were classified using Refactoring Miner [41], and fault information was obtained by extracting issues from the Jira issue tracker. After that, code quality was inspected using two tools: Technical Debt items were analyzed with SonarQube, and code smells [10] and anti-patterns [1] with Ptidej [13]. In addition, the fault-inducing and -fixing commits were identified by applying our implementation [33] of the SZZ algorithm [11].

---
[1]BetterCodeHub. https://bettercodehub.com. Accessed: May 2019.
[2]CoverityScan. https://scan.coverity.com. Accessed: May 2019
[3]SonarQube. https://www.sonarqube.org. Accessed May 2019

This dataset has been recently used by the authors for different works [4], [19], [20], [37], [34] and could be used by researchers to investigate various research questions regarding Technical Debt.

The dataset is easily accessible as we provide two alternative ways to access the data, a set of CSV files and an SQLite database file[4], in order to facilitate queries on the data.

The main contributions of this paper are:
- The *Technical Debt Dataset*. A curated set of projects where we analyzed all the commits from the time frames reported in Table 1 with four tools
- *Two data formats*. CSV files and an SQLite database, to enable researchers to efficiently query the data via SQL
- The data retrieval pipeline that produced this dataset.

The remainder of this paper is structured as follows. Section 2 describes the project selection strategy. Section 3 reports the tools used to produce the dataset, while Section 4 describes the characteristics extracted from the data. Section 5 reports on the data production pipeline. Section 6 describes the data schema. Section 7 reports on the significance of this dataset. In Section 8, we report how we plan to update the dataset, and Section 9 contains licence information. Section 10 presents the threats to validity, and in Section 11, we draw conclusions.

## 2 PROJECT SELECTION

The selected projects had to fulfill all of the following criteria:
- Developed in Java
- Older than three years
- More than 500 commits
- More than 100 classes
- Usage of Jira issue tracking systems with at least 100 issues

Moreover, as recommended by Nagappan et al. [28], we also tried to maximize diversity and representativeness by considering a comparable number of projects with respect to project age, size, and domain.

Based on these criteria, we selected 33 Java projects from the Apache Software Foundation (ASF) repository[5]. This repository includes some of the most widely used software solutions. The available projects can be considered industrial and mature, due to the strict review and inclusion process required by the ASF. Moreover, the included projects have to keep on reviewing their code and follow a strict quality process[6].

In Table 1, we report the list of the 33 projects we considered together with the number of analyzed commits, and the time frame, number of the analyzed commits, and number of reported items for each tool. The analysis time frame differs between the tools. However, all projects have been analyzed using all of the tools for the time frame of SonarQube / Ptidej.

## 3 TOOLS USED TO COLLECT THE DATA

In order to produce the dataset, we used four tools for analyzing the project data. These tools will be described in the following subsections.

### 3.1 PyDriller

PyDriller [39] is a Python framework meant for mining Git repositories. It provides easy extraction of information from a Git repository. For example, the tool supports extraction of the commit message, the number of developers, modifications, diffs, and the source code of a commit. Moreover, PyDriller calculates structural metrics of every file changed in a commit relying on Lizard[7], a tool that can analyze source code of different programming languages, both at class and method level.

### 3.2 Ptidej

Ptidej (Pattern Trace Identification, Detection and Enhancement in Java) is one of the most popular tools for detecting code smells in research [23] and has been adopted by more than 200 research works[8] Ptidej can detect a total of 18 code smells [10] and anti-patterns [1].

The tool is intended for use by research and development teams for proposing and validating ideas, methods, and tools. The goal is to improve the quality of systems implemented following the object-oriented paradigm.

This goal is achieved by introducing the concepts design, and architectural patterns. However, these are often not used (due to time pressure or programmers' unawareness) neither in the design phase nor in the coding phase. Ptidej 5 can be freely downloaded from the official online repository[9]. Since October 2014, the source code of the Ptidej Tool Suite has been open and released under the GNU Public License v2.

The Ptidej Tool Suite offers several tools for the assessment and improvement of code quality. These include such tools as PADL (Pattern and Abstract-level Description Language), which is a meta-model for describing systems at different levels of abstraction; POM (Primitives, Operations, Metrics), which a framework for computing several metrics on PADL models; and SAD, which stands for Software Architectural Defects.

SAD specifies and identifies occurrences of code smells in systems modeled using the PADL tool. SAD requires the POM module to identify code smells within the code; it employs a well-structured procedure by means of rule cards. Each time detection of a specific smell is requested, SAD imports its rule card and, employing the POM module, calculates the software metrics.

### 3.3 Refactoring Miner

Refactoring Miner [41] is an open-source tool that classifies the different refactorings in the history of Java projects. Refactoring

---

[4] SQLite Database. https://www.sqlite.org
[5] http://apache.org
[6] https://incubator.apache.org/policy/process.html
[7] Lizard GitHub Repository. https://github.com/terryyin/lizard. Accessed July 2019
[8] http://www.ptidej.net/publications/
[9] Ptidej GitHub Repository: https://github.com/ptidejteam/v5.2

**Table 1: Description of the selected projects.**

| Project Name | Git | | | Refactoring Miner | | SonarQube | | Ptidej | | Analysis Timeframe | | |
|---|---|---|---|---|---|---|---|---|---|---|---|---|
| | #Commits | #Issues | #Faults | #Commits | #Refactorings | #Commits | #Introduced TD items | #Commits | #Introduced Code Smells [10] | Git / Refactoring Miner | Jira | SonarQube / Ptidej |
| Accumulo | 10,122 | 4,744 | 2,250 | 42 | 368 | 2,641 | 1,377,049 | 0 | 0 | 10/11 - 7/19 | 10/11 - 7/19 | 10/11 - 10/13 |
| Ambari | 24,579 | 25,041 | 17,722 | 58 | 342 | 13,397 | 40,698 | 0 | 0 | 08/11 - 7/19 | 09/11 - 7/19 | 09/11 - 06/15 |
| Atlas | 2,794 | 3,284 | 1,990 | 793 | 5,029 | 2,336 | 34,997 | 0 | 0 | 11/14 - 7/19 | 05/15 - 7/19 | 12/14 - 06/18 |
| Aurora | 4,065 | 1,969 | 628 | 562 | 3,080 | 4,012 | 7,405 | 0 | 0 | 04/10 - 7/19 | 10/13 - 7/19 | 04/10 - 06/18 |
| Batik | 3,491 | 1,265 | 1,160 | 429 | 3,117 | 2,097 | 31,113 | 0 | 0 | 10/00 - 7/19 | 01/01 - 7/19 | 10/00 - 06/06 |
| Beam | 22,332 | 2,361 | 1,723 | 2,113 | 10,105 | 2,865 | 74,434 | 2,865 | 8,458 | 12/14 - 7/19 | 02/16 - 7/19 | 12/14 - 07/16 |
| Cocoon | 13,160 | 521 | 327 | 984 | 3,447 | 10,210 | 47,994 | 10,210 | 6,513 | 02/03 - 7/19 | 01/01 - 7/19 | 03/03 - 02/07 |
| Commons BCEL | 1,451 | 7,750 | 3,218 | 85 | 643 | 1,324 | 7,471 | 1,324 | 562 | 10/01 - 7/19 | 05/02 - 7/19 | 10/01 - 04/18 |
| Commons BeanUtils | 1,234 | 319 | 242 | 75 | 203 | 1,192 | 4,674 | 1,192 | 424 | 03/01 - 7/19 | 10/01 - 7/19 | 03/01 - 06/18 |
| Commons CLI | 921 | 669 | 346 | 51 | 145 | 896 | 30,300 | 896 | 3,779 | 06/02 - 7/19 | 06/02 - 7/19 | 06/02 - 02/18 |
| Commons Codec | 1,825 | 295 | 182 | 148 | 474 | 1,726 | 1,831 | 1,726 | 166 | 04/03 - 7/19 | 04/02 - 7/19 | 04/03 - 05/18 |
| Commons Collections | 3,135 | 256 | 135 | 382 | 2,663 | 2,982 | 9,566 | 2,982 | 1,175 | 04/01 - 7/19 | 05/01 - 7/19 | 04/01 - 09/18 |
| Commons Configuration | 3,077 | 108 | 73 | 488 | 1,585 | 2,895 | 4,334 | 2,895 | 869 | 12/03 - 7/19 | 01/03 - 7/19 | 12/03 - 05/18 |
| Commons Daemon | 1,087 | 297 | 190 | 7 | 11 | 980 | 371 | 980 | 13 | 09/03 - 7/19 | 08/03 - 7/19 | 09/03 - 08/11 |
| Commons DBCP | 2,010 | 598 | 284 | 174 | 600 | 1,861 | 5,390 | 1,861 | 265 | 04/01 - 7/19 | 02/02 - 7/19 | 04/01 - 06/18 |
| Commons DbUtils | 662 | 291 | 159 | 37 | 152 | 645 | 545 | 645 | 56 | 11/03 - 7/19 | 11/03 - 7/19 | 11/03 - 12/17 |
| Commons Digester | 2,145 | 305 | 149 | 136 | 405 | 2,145 | 6,336 | 2,145 | 926 | 05/01 - 7/19 | 02/01 - 7/19 | 05/01 - 08/17 |
| Commons Exec | 627 | 747 | 444 | 36 | 77 | 617 | 655 | 617 | 67 | 07/05 - 7/19 | 08/05 - 7/19 | 08/05 - 01/16 |
| Commons FileUpload | 962 | 402 | 282 | 21 | 101 | 922 | 666 | 922 | 62 | 03/02 - 7/19 | 10/02 - 7/19 | 03/02 - 10/17 |
| Commons IO | 2,180 | 540 | 368 | 140 | 494 | 2,118 | 5,381 | 2,118 | 381 | 01/02 - 7/19 | 04/03 - 7/19 | 01/02 - 06/18 |
| Commons Jelly | 1,939 | 142 | 56 | 171 | 455 | 1,939 | 6,189 | 1,939 | 764 | 02/02 - 7/19 | 07/02 - 7/19 | 02/02 - 09/17 |
| Commons JEXL | 1,655 | 191 | 119 | 344 | 1,896 | 1,551 | 33,694 | 1,551 | 1,107 | 04/02 - 7/19 | 06/03 - 7/19 | 04/02 - 05/18 |
| Commons JXPath | 598 | 455 | 265 | 66 | 396 | 597 | 4,550 | 597 | 355 | 08/01 - 7/19 | 05/02 - 7/19 | 08/01 - 11/15 |
| Commons Net | 2,117 | 719 | 438 | 91 | 325 | 2,088 | 35,565 | 2,088 | 3,738 | 04/02 - 7/19 | 02/02 - 7/19 | 04/02 - 08/17 |
| Commons OGNL | 615 | 6,056 | 3,415 | 49 | 460 | 608 | 4,483 | 608 | 362 | 05/11 - 7/19 | 11/05 - 7/19 | 05/11 - 09/13 |
| Commons Validator | 1,342 | 932 | 397 | 50 | 168 | 1,339 | 1,720 | 1,339 | 252 | 01/02 - 7/19 | 02/02 - 7/19 | 01/02 - 04/18 |
| Commons VFS | 2,288 | 133 | 84 | 293 | 1,015 | 2,067 | 3,111 | 2,067 | 549 | 07/02 - 7/19 | 03/04 - 7/19 | 07/02 - 05/18 |
| Felix | 15,427 | 194 | 147 | 1,585 | 6,610 | 596 | 10,370 | 596 | 772 | 07/05 - 7/19 | 07/05 - 7/19 | 08/05 - 10/06 |
| HttpComponents Client | 3,009 | 663 | 463 | 483 | 2,624 | 2,867 | 8,998 | 2,867 | 1,436 | 12/05 - 7/19 | 10/01 - 7/19 | 12/05 - 06/18 |
| HttpComponents Core | 3,288 | 258 | 188 | 572 | 3,435 | 1,941 | 7,531 | 1,941 | 1,255 | 02/05 - 7/19 | 07/02 - 7/19 | 02/05 - 08/17 |
| MINA SSHD | 1,787 | 566 | 285 | 512 | 3,993 | 1,370 | 7,724 | 1,370 | 1,002 | 12/08 - 7/19 | 12/08 - 7/19 | 12/08 - 06/18 |
| Santuario Java | 2,824 | 1,932 | 1,302 | 224 | 1,035 | 2,697 | 19,807 | 2,697 | 1,854 | 09/01 - 7/19 | 08/13 - 7/19 | 10/01 - 06/18 |
| ZooKeeper | 1,940 | 3,424 | 1,859 | 522 | 2,077 | 411 | 5,265 | 411 | 391 | 11/07 - 7/19 | 06/08 - 7/19 | 05/08 - 06/18 |
| **Total** | 134,812 | 67,427 | 40,890 | 11,723 | 57,530 | 77,932 | 1,840,217 | 53,449 | 37,553 | | | |

Miner takes as an input the list of commits, and returns a list of refactoring operations applied between consecutive commits.

Refactoring miner's release 1.0.0 can detect 15 types of refactorings from different types of code elements (see Table 2).

**Table 2: Refactorings detected by Refactoring Miner [41].**

| Code Element | Refactoring |
|---|---|
| Package | Change Package (move, rename, split) |
| Type | Move Class, Rename Class, Extract Superclass/Interface |
| Method | Extract Method, Inline Method Pull Up Method, Push Down Method Rename Method, Move Method Extract and Move Method |
| field | Pull Up Field, Push Down Field Move Field |

### 3.4 SonarQube

SonarQube is one of the most common open-source static code analysis tools for static quality analysis. It can be executed on premise or with the free cloud-based service on sonarcloud.io.

SonarQube calculates several metrics such as number of lines of code and code complexity, and verifies the code's compliance against a specific set of "coding rules". In case the analyzed source code violates a coding rule or if a metric is outside a predefined threshold (also named "quality gate"), SonarQube generates an "issue". The time needed to remove these issues (remediation effort) is used to calculate the remediation cost and the Technical Debt.

Each rule is classified as being related to Reliability, Maintainability, or Security of the code. Reliability rules, also named "bugs", create TD issues that "represent something wrong in the code" and that will soon be reflected in a bug. Maintainability rules or "code smells" are considered as "maintainability-related issues" in the code that decrease code readability and modifiability. It is important to note that the term "code smells" adopted in SonarQube does not refer to the commonly known code smells defined by Fowler et al. [10], but to a different set of rules. SonarQube claims that zero false-positive issues are expected from the Reliability and Maintainability rules, while Security issues may contain some false-positives[10].

---
[10]SonarQube Rules: https://docs.sonarqube.org/display/SONAR/Rules
Last Access: June 2019

SonarQube also classifies these rules into five *severity* levels[11]:
- *BLOCKER*: "Bug with a high probability to impact the behavior of the application in production", such as memory leaks and unclosed JDBC connections.
- *CRITICAL*: "Either a bug with a low probability to impact the behavior of the application in production or an issue which represents a security flaw". Examples are empty catch blocks and SQL injection.
- *MAJOR*: "Quality flaw which can highly impact the developer productivity", such as uncovered piece of code, duplicated blocks, or unused parameters.
- *MINOR*: "Quality flaw which can slightly impact the developer productivity". Examples: lines should not be too long and "switch" statements should have at least three cases.
- *INFO*: "Neither a bug nor a quality flaw, just a finding."

SonarQube also recommends immediately reviewing blocker and critical issues.

SonarQube has separate sets of rules for the most common development languages such as Java, Python, C++, and JavaScript. SonarQube version 7.5 includes more than 500 rules for Java. The complete list of rules is available online[12] but can also be found in the file "sonar_rules.csv" of the Technical Debt Dataset.

### 3.5 SZZ Algorithm

OpenSZZ [33] is our free implementation of the SZZ algorithm. The SZZ [15] algorithm tries to identify the fault-inducing commits from a project's version history. The algorithm was developed in 2005 and has since been adopted in more than 200 empirical studies [11], [6].

The algorithm is based on Git's blame/annotate feature and assumes that the fault-inducing commit of a fault is known. Usually this is done by combining data from an issue tracker and from Git's log command.

The algorithm consists of three steps. An example is provided in Figure 1.

The first step is to identify the fault-fixing commits, i.e., commits that are known to have fixed a bug. For example, in step 1 of the figure, we can see that fault AMBARI-17618 was fixed from file *Resource.java* in commit *#e8bfdb*.

The second step is to identify the changes that fixed the fault. The algorithm inspects the files that were changed in the fault-fixing commit. From these files, the algorithm identifies the changes related to updating a data structure, i.e., bug-fixing activities. In the example, step 2 shows the differences between commit *#e8bfdb* and its predecessor (*#300a7e*) in the *Resource.java* file. In this case, in order to fix the bug, the data structure at line 188 was changed. Therefore, SZZ identifies the changes that introduced bug *AMBARI-17618* through the history of the source configuration management system (GitHub).

The last step determines in which commit the code that caused the fault was introduced. This is done by using Git's annotate/blame tool. In the example, step 3 shows commit *#a2d7c9* being flagged as a potential bug-introducing change by SZZ.

---

[11]Severity of SonarQube Issues and Rules:'https://docs.sonarqube.org/display/SONAR/Issues Last Access: June 2019
[12]https://rules.sonarsource.com/java Last Access: June 2019

Each commit is tagged with the information about the retrieved fault from the SZZ algorithm. Commits can be of one of three fault types: inducing (I), fixing (F), or not related to faults (N).

It is important to notice that we applied the SZZ algorithm as is, without applying any filtering. It could be possible to apply different filters, considering different contextual information provided in the dataset.

## 4 COLLECTED MEASURES

The Technical Debt Dataset includes information on the following characteristics:

- *Commit Information*. Information about each commit obtainable through Git's log command. The information includes, for example, the commit hash, date, and message.
- *Refactorings* List of refactorings applied in each commit, obtained from Refactoring Miner.
- *Code Quality*. List of detected issues related to the code quality of a commit. These consist of style violations and other low-level technical issues monitored by SonarQube, and the detected anti-patterns [1] and code smells [10] such as Large Class and Long Method.
- *Jira issues*. List of all detected issues from the project's issue tracker.
- *Fault-Inducing and Fault-Fixing Commits*. For each fixed fault in a project, the commit where the fault was created (fault-inducing) and the commit where the fault was fixed (fault-fixing) are determined.

In the following subsections, we will introduce the different characteristics in more depth and describe how we extracted them from the data.

### 4.1 Commit and Changes Information

We extracted the commit information from the Git repositories using PyDriller [39]. This tool was used instead of connecting directly to the GitHub API for maintainability reasons. Connecting to the GitHub API or parsing the Git Logs would require developing a custom script, whereas PyDriller currently allows extracting the same information without major effort.

Thanks to PyDriller, we saved all the information available in the GitLog, including information on each file modified in each commit. Moreover, we collected information on the refactoring applied in each commit by executing Refactoring Miner.

### 4.2 Code Quality

As in the case of software quality, we collected three sets of information:

- **Software Metrics**. 30 different software metrics, including:
  - size-related metrics, such as number of classes, number of packages, and number of lines of code
  - complexity metrics, such as Cyclomatic Complexity [27] and cognitive complexity [3]
  - test coverage, including test coverage as well as the number of lines not covered by tests
  - duplications, including number of duplicated lines and duplicated files.

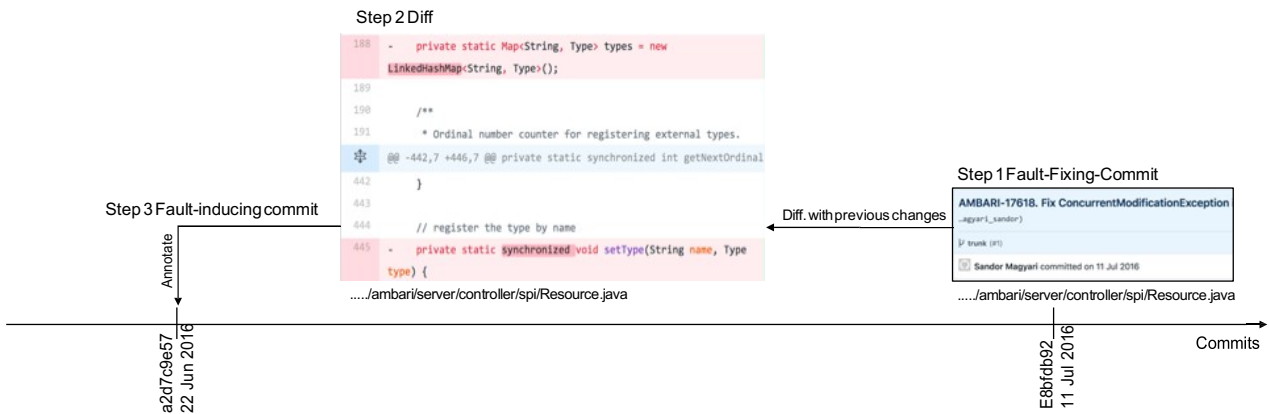

**Figure 1: An example of the execution of the SZZ algorithm.**

- **Technical Debt** information detected by SonarQube, including:
  - Issues detected in the projects, classified as "bugs", "code smells", and "security vulnerabilities" (see Section 3.4)
  - Technical Debt remediation time, including the time for refactoring "bugs", SonarQube "code smells", and "security vulnerabilities" issues.
- **Code Smells** [10] **and Anti-Patterns** [1] detected by Ptidej.

The complete list of metrics collected in the dataset together with their description is reported in Table 3, while the collected anti-patterns and code smells are described in Table 4. For reasons of space, the list of SonarQube issues and rules is not reported here, but it is available in the dataset.

### 4.3 Jira issues

For each project, we extracted all issues from the Jira issue tracker. The information includes, for example, the creation and resolution date of the issue, issue key, and the type and priority of the issue. Descriptions for the extracted fields can be found from Jira's documentations[13].

### 4.4 Fault-Inducing and Fault-Fixing Commits

For this step, we first extracted the faults from Jira issues. This was possible as the ASF policy states that all faults reported in the issue tracker have to be tagged as "Bug".

Faults are commonly discovered after the code has already been committed. Therefore, in order to ensure that we would find the vast majority of faults related to commits, we analyzed the projects from their first commit until the end of 2015, considering all the faults raised until the end of March 2018. We believe that the vast majority of the faults introduced in the commits from the time frames reported in Table 1 should have been discovered after more than two years.

We tried to identify the fault-inducing and -fixing commits for all the faults retrieved from Jira. This was done by applying the SZZ algorithm [15] to the selected projects. We used our own, publicly available implementation of the algorithm [33] for the analysis.

As stated in subsection 3.5, the algorithm assumes that the fault-fixing commit is known for every fault. The selected ASF projects fulfill this assumption, as the ASF policy requires that in all fault-fixing commits the developer must report the fault-id from the issue tracker in the commit message. Therefore, we relied on this mechanism to map the commits from which developers had removed faults.

## 5 DATASET PRODUCTION

We performed the following tasks for each project independently. We first cloned the repository. Then we iterated each commit using PyDriller [39].

For each commit, we performed the following steps:

- Retrieval of the commit information from the GitLog using PyDriller
- Classification of the refactorings using Refactoring Miner
- Analysis of the code with SonarQube using the default quality model (Sonar way)
- Analysis of the code with Ptidej

This set of tasks was the most time-consuming part, since each commit took an average of four minutes to be analyzed with SonarQube and 50 seconds with Ptidej. As the open-source license of SonarQube does not allow analyzing multiple projects in parallel, we analyzed the projects sequentially. However, sequentiality was also required because SonarQube calculates evolutionary metrics based on the changes performed in each commit. SonarQube and Ptidej were executed in parallel.

These three steps took a total of 220 days on a Ubuntu server with 128GB RAM and 50 cores. However, because of the limitation of SonarQube's open-source license, the SonarQube analysis used only one core and an average of 12GB RAM, while Ptidej used one core and an additional 6GB RAM on average. It is important to note that not all the commits were analyzed correctly, as SonarQube or Ptidej raised exceptions when a commit did not compile properly. However, we stored the information about all the commits in the COMMITS table, while storing only the correctly performed analyses in the other tables.

---
[13] https://confluence.atlassian.com/jiracorecloud/advanced-searching-fields-reference-765593716.html

While the script analyzing the commits was running, we extracted the fault information from Jira by means of the Jira APIs. The dataset includes all the issues with the status "closed". This task took less than two days of execution time on a virtual machine with 4 cores and 8GB RAM.

At the end of the execution of both scripts, we executed our SZZ implementation [33] to identify the fault-inducing commits. This task took ten days of execution time on the same machine used for the commit analysis (SonarQube and all the other services were stopped in order to enable SZZ to use as many resources as possible).

After the extraction of all the data, we performed a data-sanity check to evaluate possible inconsistencies, including duplicated and unexpected values.

Finally, we imported the data into an SQLite database to facilitate queries on the data.

## 6 THE DATA SCHEMA

The Entity Relationship schema of the dataset is presented in Figure 2. Regarding the fields in the tables, the fields needed to distinguish a row are underlined and "..." means that the table has more fields than presented in the figure. Looking at the fields of the tables, it is evident that the database has not been normalized. However, this is intentional and data is often replicated in different tables to speed up queries by avoiding heavy use of joins between tables.

The tables are briefly described below and the data of each a table can be found in the respective identically named CSV file in the Technical Debt Dataset.

- Table PROJECTS contains the links to the GitHub repository and the Jira issue tracker of each project.
- Table SONAR_MEASURES contains the different measures SonarQube analyses from the commits. They contain, for example, number of code lines in the commit, the code complexity, and number of functions. Note, that the table contains information about the code in the commit rather than the commit itself.
- Table COMMITS reports the commit information retrieved from the git log, including the commit hash, the commit message, the author and committer name, date and timezone, the list of branches that contain the commit, in_main_branch (true if the commit is in the main branch), merge (true if the commit is a merge commit), parents (list of the commit parents).
- Table COMMITS_CHANGES contains the changes performed in each commit, including the old path of the file (can be _None_ if the file is added), the new path of the file (can be _None_ if the file is deleted), the type of the change reported in the git Log (Added, Deleted, Modified, or Renamed), the diff of the file as Git presents it, the number of lines added, the number of lines removed, the number of lines of code of the file (nloc), the Cyclomatic Complexity of the file (calculated with Lizard), the number of tokens of the file, the list of methods of the file (might be empty if the programming language is not supported or if the file is not a source code file). .

- Table JIRA_ISSUES contains Jira issues for the analyzed projects. Fields include the key (the issue-id used by Jira to identify each issue), the creation and resolution dates, the priority assigned by the developers, and many others.
- Table FAULT_INDUCING_COMMITS reports the results from the execution of the SZZ algorithm. Each row in the table is a fault retrieved from Jira. The fields include information on the fault-inducing commits (*FaultInducingCommitHash* and *FaultInducingTimestamp*) and on the fault-fixing commits (*FaultFixingCommitHash* and *FaultFixingTimestamp*). If the fault-fixing commit could not be determined, the related fields are left empty.
- Table REFACTORING_MINER reports the list of refactoring activities applied in the repositories. The table contains the project, commit hash, the type of refactoring applied, and the details of the refactoring activity.
- Table SONAR_ISSUES lists all of the SonarQube issues as well as the anti-patterns and code smells detected by Ptidej. The value of field *squid* of the issues detected by SonarQube starts with either the prefix squid: or common-java:, while the code smells and anti-patterns detected by Ptidej are identified using the prefix code_smells:. Each row represents one issue, reporting which rule, code smell, or anti-pattern was violated, where it was found, when it was created, and the fixing date if it has been fixed. In addition to these features, several other attributes about the issue are reported as well.
- Table SONAR_RULES lists the rules monitored by SonarQube. For each rule, the feature *description* provides a short description. In addition, the table contains, for example, the type and severity assigned by SonarQube for the rule.

### 6.1 How to use the dataset

The dataset is accessible online [36] and it is provided both as a set of CSV files and as an SQLite database. In both cases, the same data is provided; it is structured as shown in Figure 2.

Each CSV file contains the data of one table. The SQLite[14] database is provided as a .db file and can be opened with a SQLite browser such as SQLiteStudio[15]. After opening, the data can be queried using SQL by selecting "Tools" and "Open SQL Editor". Data can also be exported as a CSV file by selecting "Tools" and "Export" and then exporting a whole table or an SQL query.

Listing 1 provides an example on how to use data from different tables. The example combines data from tables SONAR_MEASURES and SONAR_ISSUES in order to get the number of SonarQube issues in a project.

## 7 SIGNIFICANCE

Source code analysis has been attracting more attention from research and industry, mainly because of the availability of tools like SonarQube. More than 100K companies are currently using SonarQube and are therefore considering the information, metrics, and issues reported by it.

---

[14] https://www.sqlite.org/index.html
[15] SQLiteStudio. https://sqlitestudio.pl

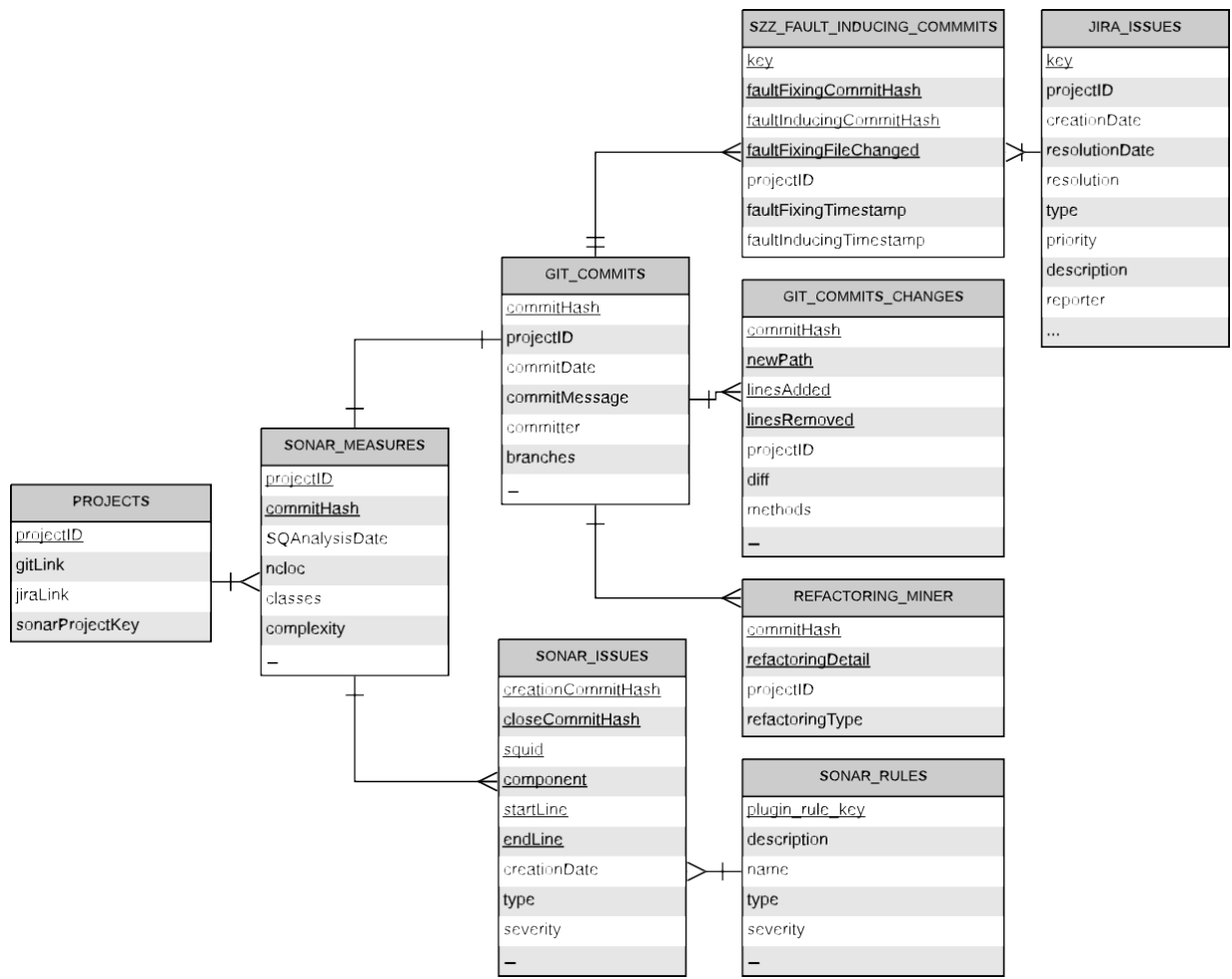

Figure 2: Entity Relationship Diagram of the dataset.

```sql
SELECT  SONAR_ISSUES.projectID,  count(SONAR_ISSUES.projectID)  AS  numberOfBugIssues, numberOfVulnerabilityIssues
FROM SONAR_ISSUES
JOIN
   (SELECT  SONAR_ISSUES.projectID,  count(SONAR_ISSUES.projectID)  AS  numberOfVulnerabilityIssues
    FROM SONAR_ISSUES
    WHERE  TYPE='VULNERABILITY'
    GROUP BY  SONAR_ISSUES.projectID)  AS  V  ON  SONAR_ISSUES.projectID  =  V.projectID
WHERE TYPE='BUG'
GROUP BY  SONAR_ISSUES.projectID
```

Listing 1: An SQL query for getting the number of "Bug" and "Vulnerability" SonarQube issues in a project.

```sql
SELECT   GIT_COMMITS.commitMessage,  GIT_COMMITS.author
FROM     SZZ_FAULT_INDUCING_COMMITS
         JOIN  GIT_COMMITS
           ON  SZZ_FAULT_INDUCING_COMMITS.faultFixingCommitHash  =  GIT_COMMITS.commitHash
```

Listing 2: An SQL query for getting the commit message and author of all the commits fixing a SonaerQube issue.

As for the selected projects, a number of studies have been carried out about projects in the Apache Software Foundation ecosystems [9], [8], [2], [31], but they mainly focus on different languages or on different types of issues.

The areas that can benefit from our Technical Debt Dataset are statistical machine learning, natural language processing on source code, but also empirical studies on Technical Debt.

Possible future work could also be related to the investigation of the harmfulness of the different SonarQube issues, including their fault- and change-proneness. This area has already been widely investigated considering code smells [14], [17], [38], [31] but researchers have never considered TD issues detected by SonarQube from the point of view of fault- and change-proneness.

Our dataset will allow researchers to compare the effect of code smells with those of SonarQube violations, as the data include the analysis of code smells with Ptidej and the SonarQube issues on the same projects.

Different research questions could be investigated with the Technical Debt Dataset. Besides the possibility to replicate the existing works on code smells, it would be possible to investigate many other research questions, such as:

- How are the projects maintained?
- How are faults and Technical Debt related?
- How can we detect buggy commits?
- How do different Technical Debt issues co-evolve?
- Does the evolution of Technical Debt issues follow patterns?
- Do these patterns differ between projects?
- Which refactoring activity introduce more technical debt?
- Which Technical Debt issue is more prone to refactoring?

## 8   UPDATE

Open-source projects are continuously evolving and thus our dataset will also need to be updated regularly. We provide a GitHub repository to allow other researchers to contribute by analyzing other projects [36].

We will provide new releases of the dataset whenever a new project has been analyzed completely.

## 9   LICENSE

The *Technical Debt Dataset* has been developed only for research purposes. It includes the historical analysis of each public repository, including commit messages, timestamps, author names, and email addresses. Information from GitHub is stored in accordance with GitHub Terms of Service (GHTS), which explicitly allow extracting and redistributing public information for research purposes [16].

The *Technical Debt Dataset* is licensed under a Creative Commons Attribution-NonCommercial-ShareAlike 4.0 International license.

## 10   THREATS TO VALIDITY

Researchers should always consider potential threats to the validity of their research.

Regarding the data collection process and the traditional trade-off between being up-to-date and curation [6], we preferred emphasizing the curation of the dataset rather than providing a limited up-to-date amount of data. Moreover, we relied on PyDriller to iterate over the different commits and to retrieve the commit information. We are aware that both Ptidej and SonarQube might analyze the code incorrectly under some conditions. Moreover, regarding SonarQube, we adopted the out-of-the-box "Sonar way" model. Although SonarQube recommends customizing the set of rules used [17], practitioners are reluctant to customize it and commonly rely on the standard rule set [42]. Querying the SonarQube public instance APIs [18] reveals that more than 98% of the public projects (14,732 projects out of 14,957) use the "Sonar way" rule set. Moreover, it is important to notice that the SonarQube remediation time can be overesttimated [35]. Therefore, future studies consiering the remediation time should consider this threat.

Another possible tool-related threat concerns our implementation of SZZ. The implementation of the SZZ algorithm has been manually validated in [33]. However, we are aware that in some cases, the SZZ algorithm might not have identified fault-inducing commits correctly because of the limitations of the line-based diff provided by Git, and also because in some cases bugs can be fixed by modifying code in another location than the lines that induced them. Moreover, we rely on the developers that must report the ID of the fault to the commits that fixed it to identify the fault-fixing commit, but we are aware that, even if the ASF is enforcing this approach, some faults could be not correctly labeled.

Another important threat is related to the generalization of the dataset. We selected the list of projects based on different criteria (see Section 2). However, even though the projects are widely used in industry, they cannot possibly represent the whole open-source ecosystem. Moreover, since the dataset does not include industry projects, we cannot make any speculation on closed-source projects.

## 11   CONCLUSION

In this paper, we presented the Technical Debt Dataset [36]. It is the largest source code dataset analyzing Java projects with different tools, including tools widely used in industry and research.

We described the pipeline we adopted to collect the data. The Technical Debt Dataset is made available both as an SQLite database file and as CSV files in order to facilitate queries on the data. The metadata is carefully provided as part of the download.

The creation of the dataset took more than 200 days due to the license limitation of SonarQube, whose open-source license does not allow running analyses in parallel.

The results will allow researchers to perform various studies on a common dataset, comparing their results and concentrating more on their research goals instead of on data collection. The current research on code smells could be extended in the future to SonarQube issues, enabling researchers to investigate different quality aspects, including technical debt, maintenance effort, fault-proneness of different metrics, code smells, and several other aspects.

We are planning to extend this dataset in the future applying different measurement tools to the same projects. Moreover, we are planning to investigate further datasets such as [30] that could be used to complement the data of our dataset. Moreover, we are planning to extend our previous work with this version of the dataset, applying NLP on requirements and Jira issues [18], and analyzing the relationship between technical debt and coupling [4]. Moreover we are planning to analyze differences between technical debt issues in open source projects and in SMEs [21] and to use the results as baseline for continuous quality monitoring approaches [16], [24].

---

[16] GitHub Terms of Service. goo.gl/yeZh1E Accessed: May 2019

[17] SonarQube Quality Profiles: https://docs.sonarqube.org/display/SONAR/Quality+Profiles Last Access:June 2019
[18]　https://docs.sonarqube.org/display/DEV/API+Basics

---

[18] https://docs.sonarqube.org/latest/user-guide/metric-definitions/

**Table 3: The software metrics collected in the dataset. Complete descriptions can be found in the SonarQube documentation**

| Metric | Description |
|---|---|
| **Size** | |
| Number of classes | Number of classes (including nested classes, interfaces, enums and annotations). |
| Number of files | Number of files. |
| Lines | Number of physical lines (number of carriage returns). |
| Ncloc | Also known as Effective Lines of Code (eLOC). Number of physical lines that contain at least one character which is neither a whitespace nor a tabulation nor part of a comment. |
| Ncloc language distribution | Non Commenting Lines of Code Distributed By Language |
| Number of classes and interfaces | Number of Java classes and Java interfaces |
| Missing package info | Missing package-info.java file (used to generate package-level documentation) |
| Package | Number of packages |
| Statements | Number of statements. |
| Number of directories | Number of directories in the project, also including directories not containing code (e.g., images, other files...). |
| Number of functions | Number of functions. Depending on the language, a function is either a function or a method or a paragraph. |
| Number of comment lines | "Number of lines containing either comment or commented-out code. Non-significant comment lines (empty comment lines, comment lines containing only special characters, etc.) do not increase the number of comment lines." |
| Number of comment lines density | Density of comment lines = Comment lines / (Lines of code + Comment lines) * 100 |
| **Complexity** | |
| Complexity | It is the Cyclomatic Complexity calculated based on the number of paths through the code. Whenever the control flow of a function splits, the complexity counter gets incremented by one. Each function has a minimum complexity of 1. This calculation varies slightly by language because keywords and functionalities do. |
| Class complexity | Complexity average by class |
| Function complexity | Complexity average by method |
| Function complexity distribution | Distribution of method complexity |
| File complexity distribution | Distribution of complexity per class |
| Cognitive complexity | How hard it is to understand the code's control flow. |
| Package dependency cycles | Number of package dependency cycles |
| **Test coverage** | |
| Coverage | It is a mix of Line coverage and Condition coverage. Its goal is to provide an even more accurate answer to the following question: How much of the source code has been covered by the unit tests? |
| Lines to cover | Number of lines of code which could be covered by unit tests (for example, blank lines or full comments lines are not considered as lines to cover). |
| Line coverage | On a given line of code, Line coverage simply answers the following question: Has this line of code been executed during the execution of the unit tests? |
| Uncovered lines | Number of lines of code which are not covered by unit tests. |
| **Duplication** | |
| Duplicated lines | Number of lines involved in duplications |
| Duplicated blocks | Number of duplicated blocks of lines. |
| Duplicated files | Number of files involved in duplications. |
| Duplicated lines density | = (duplicated lines ÷ lines) * 100 |

**Table 4: Code smells and anti-patterns detected in our projects. Descriptions are adapted from [10] and [1].**

| Name | Description |
|---|---|
| Duplicated code | The same code reused in different locations. |
| Blob | A big class (usually a singleton) that has dependencies with data contained in other data classes. It could monopolize several system operations. |
| Class data should be private | Class publicly exposing variables. |
| Cyclomatic complexity | Also referred to as McCabe's Cyclomatic Complexity. Refers to methods that should be simplified since they have too many independent execution paths. |
| Downcasting | A cast to a derived class. |
| Excessive use of literals | Too many literal variables embedded into the code instead of being declared as constants. |
| Feature envy | An object gets at the fields of another object to perform some sort of computation or make a decision, rather than asking the object to do the computation itself. |
| Functional decomposition | A class with too many functionalities, which needs to be broken down into smaller and simpler classes. |
| God Class | A huge class implementing different responsibilities. |
| Inappropriate intimacy | A method that has too much intimate knowledge of the inner workings, inner data... of another class or method. |
| Large class | A class that is too big. |
| Lazy class/Freeloader | A class with very limited functionalities. |
| Orphan variable or constant class | A class containing variables used in other classes. |
| Refused bequest | A children class that never uses the inherited methods. |
| Spaghetti code | A class with a very complex control flow. |
| Speculative generality | An abstract class with a very limited number of children who are not using its methods. |
| Swiss army knife | A class that is providing many services for different purposes, such as a Swiss army knife. |
| Tradition breaker | A class that, despite inheriting from another class, does not fully extend the parent class and has no subclasses. |
| Excessively long identifiers | Variable names that re too long and do not respect the naming conventions. |
| Excessively short identifiers | Variable names that are too short and do not allow understanding the purpose of the variable. |
| Excessive return of data | A method that returns more than what is needed from the calling methods. |
| Long method | A method that is too long. |
| Too many parameters | Methods with too many parameters, which usually become hard to read and to test. |